\documentstyle[galley,graphicx,epsf]{mn}

\begin{document}

\title {Lithium Abundances  of the Local Thin Disk Stars}

\author[Lambert \& Reddy] {David L. Lambert$^{1}$ and Bacham E. Reddy$^{1,2}$ \\
$^{1}$ Department of Astronomy, University of Texas, Austin, Texas 78712\\
$^{2}$ Indian Institute of Astrophysics (CREST campus, Hosakote), Koramangla,
Bangalore-560034}


\maketitle

\label{firstpage}

\begin{abstract}

Lithium abundances are presented for a sample of 181 nearby F and G dwarfs
with accurate {\it Hipparcos} parallaxes.
The stars 
are on  circular orbits about the Galactic
centre and, hence, are identified as belonging to the thin disk.
 This sample is combined
with two published surveys 
to provide a catalogue of lithium
abundances, metallicities ([Fe/H]), masses, and ages for
451 F-G dwarfs, almost all belonging to the thin disk. The lithium
abundances are compared and contrasted with published 
lithium abundances for F and G stars in local open clusters. The
field stars span a larger range in [Fe/H] than the clusters
for which [Fe/H] $\simeq 0.0\pm0.2$. The initial (i.e., interstellar)
lithium abundance of the solar neighborhood, as derived from stars
for which astration of lithium is believed to be unimportant, 
is traced from $\log\epsilon$(Li) = 2.2 at [Fe/H] = $-1$ to
$\log\epsilon$(Li) = 3.2 at  $+0.1$.
This form for the evolution is dependent on the assumption that astration of lithium
is negligible for the stars defining the relation. An argument is
advanced that this latter assumption may not be entirely correct,
and, the evolution of lithium with [Fe/H] may be flatter than previously
supposed.
A sharp Hyades-like Li-dip is not seen among the field stars and appears
to be replaced by a large spread 
 among lithium abundances  of stars more massive than
the lower mass limit of the dip.  
Astration of lithium by stars of masses too
low to participate in the Li-dip is discussed. These stars show little to
no spread in lithium abundance at a given [Fe/H] and mass.

\end{abstract}


\section{Introduction}
Astrophysicists continue to find excitement in studying -- observationally
and theoretically -- stellar abundances of lithium. Lithium
abundances offer  insights into a wide variety
of problems, principally those related to the
nucleosynthesis of lithium at an assortment of sites and to the astration of
lithium by stars.
The  aim of our survey of lithium abundances in main sequence
F-G stars  was to document the astration 
of lithium including  as a function of
stellar mass, age, and metallicity.
We  do this by assembling lithium
abundances for a sample of 451
F and G disk stars including
157  
stars for which lithium abundances are presented
for the first time.  We 
 compare and contrast the lithium
abundances of these field stars with the abundances
reported for similar stars in local open clusters.
Our sample complements the clusters in that it
extends to lower metallicity.
 It is not
our intent to confront the abundances with theoretical
projections and predictions. 


Throughout the paper, we assume that the observed Li\,{\sc i}
6707\AA\ resonance doublet and the derived lithium abundance
refer to the isotope $^7$Li with a negligible
contribution from the isotope $^6$Li. There are two strong
reasons for our assumption. First, all but one proposed
mode of lithium synthesis makes $^7$Li with no or 
negligible coproduction of $^6$Li. Coproduction of both Li isotopes 
occurs in the interstellar medium
between relativistic cosmic rays and ambient nuclei, as in collisions
between protons and oxygen nuclei providing $^6$Li, $^7$Li as well as
$^9$Be,  $^{10}$B, and $^{11}$B as fragments of the $^{16}$O
nucleus. Yields of $^6$Li and $^7$Li by this
process may be assessed from measurements of $^9$Be
abundances for which spallation reactions are considered to be
the sole mode of synthesis. This assessment shows the
$^6$Li abundance expected of disk main sequence stars to
be negligible. Second, astration  of atmospheric lithium by
mixing with the interior is driven by proton capture with 
a capture  cross-section  for $^6$Li that is about a factor of
80 larger than the cross-section for $^7$Li.  Thus, except in
very special or contrived circumstances even mild loss of surface lithium
by mixing and proton capture leads to complete astration of $^6$Li. Note that
not all proposed theories of the Li-dip (or general astration)
 invoke destruction of
surface lithium. Theories ascribing the Li-dip to a diffusion
of Li out of the atmosphere may predict an
alteration of the isotopic ratio. Even in these cases, the
lithium abundance is probably dominated by $^7$Li.

\section{Stellar samples and lithium abundances}

Our investigation is based on
lithium abundances of  451 F and G nearby main sequence
 stars 
drawn from three surveys.
Table 1 provides the HD number  and
basic information for each star.
 The  majority of the  sample has evolved
off the zero-age main sequence. This condition reflects the selection
criteria used by the surveyors. For stars included in two or more
of the surveys, we have adopted a mean lithium abundance, and
mean estimates for the important atmospheric parameters including the
effective temperature and the metallicity.

 With the exception of a
few stars from Chen et al.'s (2001) sample, all of the stars
belong to the local thin disk, i.e., they move about  the Galactic
centre in a roughly circular orbit at the Sun's Galactocentric
distance. The Galactic thick disk and halo are very
poorly represented in the combined sample. 
Thin disk stars of the same metallicity ([Fe/H]) have the
same chemical composition to within a narrow range, i.e., [X/Fe]
is the same for stars of the same [Fe/H], and [X/Fe] changes
only slightly with [Fe/H] (Reddy et al. 2003).
The three surveys excluded spectroscopic binaries where previously
known or detected in the course of the survey, but a few surely
remain undetected. It is unlikely, however, that Table 1
includes tidally-locked binaries which may defeat the usual
processes of astration of lithium.

\subsection{Reddy et al. (2003)}

Reddy et al. (2003) reported  abundance analyses of
181 F and G stars based on high-resolution high signal-to-noise
ratio optical spectra and model atmospheres.
These field stars span the temperature range  5550 $\leq$ $T_{\rm eff}$
$\leq$ 6500 K and metallicities $-$0.80 $\leq$ [Fe/H] $\leq$
$+$0.20 with an emphasis on the sub-solar metallicities.
These stars, all belonging to the Galactic thin disk, were largely chosen to
lie off the zero-age main sequence in order that an evolutionary age could be
determined.

  Lithium was
not among the 27 elements considered. For this paper, we
extended the analysis to lithium using the 6707\AA\
Li\,{\sc i} resonance doublet. Selection of models and atmospheric
parameters are discussed by Reddy et al. Basic atomic data
for the 6707 \AA\ doublet are taken from Reddy et al. (2002).
The LTE Li abundances were corrected for non-LTE effects using
Carlsson et al.'s (1994) recipe; the corrections are small,
attaining only 0.10 dex in the most extreme cases. 
Lithium abundances obtained for 137 of the 181 stars and upper limits
to the abundance for the other 44 stars
are expected to be accurate to about
$\pm$0.1 dex with the uncertainty ($\pm$ 100 K) in the
effective temperature as a leading source of error. The
$\pm$0.1 dex is
negligible with respect to the 2~dex or more depth of the Li-dip,
small with respect to the intrinsic scatter outside the dip, and
much less than the apparent 1~dex  growth of the 
lithium abundance from [Fe/H] = $-$1 to [Fe/H] = 0.

\subsection{Chen et al. (2001)}

Chen et al. surveyed the lithium abundance in 185 main sequence
stars sampling the interval 5600 $\leq$ $T_{\rm eff}$ $\leq$ 6600 K
and $-$1.4 $\leq$ [Fe/H] $\leq$ +0.2. A major fraction,
133 of 185 stars, were drawn from an earlier paper: Chen et al. (2000)
had given abundances for a mix of elements but not including lithium.
The remaining fraction came from a reanalysis 
of stars analysed by Lambert, Heath \& Edvardsson (1991).
The survey's stars lie off the zero-age main sequence so that
evolutionary ages may be estimated from stellar evolutionary
tracks.

The method of abundance analysis chosen by Chen et al. (2000)
and Chen et al. (2001) 
resembles  ours and included  non-LTE
corrections  from Carlsson et al. (1994).  Reddy et al.
(2003) compare elemental abundances for stars in common with Chen et al.
(2000) and find very good agreement, also for the derived atmospheric
parameters including the effective temperature and the metallicity.
 This agreement extends to
the lithium abundances for  24 stars in common with Chen
et al. (2001). The
mean difference in abundance (us minus them) is a mere
0.03$\pm$0.04 dex. For four stars, the difference in abundances
is quite large, that is 0.2 to  0.6 dex, and  traceable to
differences in the equivalent width of the 6707\AA\ line.
In particular, the 0.6 dex difference arises for a star with a
very weak lithium line.
Given this level of agreement, we merge the two samples
without  adjustments
to the lithium abundances,  the effective temperatures,  and metallicities.
For stars in common, we adopt the average of the two
lithium abundances, effective temperatures, and [Fe/H]
determinations.

\subsection{Balachandran (1990)}

Balachandran (1990) determined lithium abundances for nearly
200 field F stars. She sampled the temperature range 
7000 $\leq T_{\rm eff}$ $\leq 6000$ K with most stars having
metallicities in the range [Fe/H] from $-$0.6 to +0.2. In
contrast to the preceding three samples which were restricted
to sharp-lined stars, Balachandran explicitly included
broad-lined stars in order to investigate the effect of
rotation on lithium surface depletions.
Her sample  comprised stars on and  off the 
zero age main sequence. 

Four of Balachandran's stars
 are in our sample. 
Judged by these common stars, there is
no significant difference in the lithium abundances from
Balachandran and ourselves: the mean difference (us minus Balachandran)
is  -0.1$\pm$0.06 dex. 
 There are 10 stars that are common to Chen et al. and Balachandran.
The mean differences (Chen minus Balachandran) in atmospheric
parameters, and the Li abundances are very small: $\Delta$$T_{\rm eff}$ = $-$27 $\pm$76 ~K,
$\Delta$log $g$ = $-$0.13$\pm$0.08, $\Delta$[Fe/H] = 0.08$\pm$0.08, 
and $\Delta$log $\epsilon$(Li) = $-$0.04$\pm$0.09. These comparisons 
show no systematic differences either
in atmospheric parameters or in Li abundances between the three
samples. Thus, we have simply
adopted Balachandran results, after correction   for
the  small non-LTE effects.

\begin{table*}
\centering
\begin{minipage}{110mm}
\caption{The catalogue.
{\bf  Sample lines of the table are shown here.
The complete table  is available electronically. }}
\begin{tabular}{rccrrrrrr}
\hline \hline
HD   &   [Fe/H]   &  $T_{\rm eff}$   &  log $g$   & $M_{\rm v}$   &  mass   &  Age   &  log $\epsilon$(Li) & Source \\
     &            &   (K)            &            &            &  (M$_{\odot}$) & (Gyrs) &        &  \\
\hline
101 &   -0.29  &   5826  &    4.36   &   4.55    & 0.88  &  12.5   &   2.21  &    LR \\
153  &  -0.11  &   5791  &    3.80   &   2.89    &  1.32  &  3.44   & $<$0.69  &    LR \\
330  &  -0.27  &   5775  &    3.84   &   3.10    &  1.19  &  4.60   &   2.48  &    LR \\
400  &  -0.32  &   6096  &    4.16   &   3.61    &  1.05  &  6.46   &   2.25  &    CB \\
693  &  -0.48  &   6132  &    4.12   &   3.51    &  1.04  &  6.11   &   2.39  &    CB \\
912  &  -0.26  &   6011  &    3.82   &   2.92    &  1.22  &  4.12   &   $<$1.01  &    LR \\
1671 &  -0.09 &    6471  &    3.84  &    3.51   &   1.25 &   2.59  &    2.88 &    B \\
2454 &   -0.37 &    6418  &    4.09  &    3.26   &   1.20 &   3.50  &   $<$1.60 &     LR\\
2630 &   -0.17 &    6685  &    4.17  &    2.21   &   1.58 &   1.66  &    2.83 &     B \\
... &    ...   &     ...   &    ...   &    ...    &  ...    &   ...  &   ...  &   ...\\

\hline \\
\end{tabular}
\raggedright
Note:- LR: Lambert \& Reddy (this study); B: Balachandran (1990); C: Chen et al. (2001);
LRB, LRC or BC: identify stars in common to two surveys.
\end{minipage}
\end{table*}

\section{Stellar luminosities, masses, and ages}

Interpretation of the lithium abundances in terms of
Galactic chemical evolution and stellar astration is helped greatly by
examining the abundances as a function of stellar
mass, composition, and age. The composition (here, [Fe/H]) is taken
from the surveys.  To determine mass and age, we compare the  stars in
Hertzsprung-Russell (H-R) diagrams against a  set of theoretical stellar
evolutionary tracks for the same metallicity.
 This comparison calls for the absolute visual
magnitude $M_{\rm V}$, effective temperature $T_{\rm eff}$, and
metallicity [Fe/H] of each star. As our choice of evolutionary
tracks, we use the set provided by Girardi et al. (2000).

The $M_{\rm V}$ for a star in our or Balachandran's sample
is computed using the 
parallax and the apparent magnitude  listed in the {\it Hipparcos} catalogue. 
For Chen et al.'s (2001) stars, we adopt
their $M_{\rm V}$, also computed from the  {\it Hipparcos} parallax.  
Interstellar extinction is ignored for these stars which are
within 150 pc of the Sun. As noted above, the published $T_{\rm eff}$
and [Fe/H] are adopted.

Given $M_{\rm V}$, $T_{\rm eff}$, and [Fe/H] for a star,
its mass and age were estimated by interpolation among the
evolutionary tracks provided by Girardi et al. at 0.01$M_\odot$
steps in mass and 0.01 dex steps in [Fe/H]. 
Chen et al. (2000) give masses and ages for their stars derived from
evolutionary tracks computed by VandenBerg et al. (2002). We have
rederived these quantities using the Girardi et al. tracks and
find only slight 
differences. 
The interpolation
is accurate to 0.005 $M_\odot$ in mass almost independently
of age and [Fe/H] provided that the star does not lie near the hook
in the tracks.  
In the case of the few stars that fall near the hook 
in the evolutionary tracks, there is an ambiguity
of about 0.1$M_\odot$.
We assign higher weight to the track region where evolutionary
rate is slower.

For most of the sample, the uncertainty in $T_{\rm eff}$ translates
to an error in age but not a significant error in mass; a star evolves
approximately at constant $M_V$. An error in [Fe/H] of about 0.2 dex
translates to one in the derived mass of about 0.05$M_\odot$. The
uncertainties in the measured [Fe/H] should be much smaller than
this illustrative estimate and, hence, the associated error in the
derived mass may be ignored. The principal effect of an error in $M_V$
is an error in the derived mass. On the assumption that the
parallax is the sole source of uncertainty: $\Delta M_V = 2.2\Delta\pi/\pi$.
In our sample, 448 stars or 95$\%$ have a parallax accurate to
10\% or better or $\Delta M_V < 0.22$ which translates to a mass 
uncertainty of about 0.05$M_\odot$. Our acceptance limit for a parallax
was $\Delta\pi/\pi = 0.2$ which corresponds to a mass uncertainty of
about 0.1$M_\odot$.

\section{Setting the stage with open clusters}

By way of an introduction to the discussion of the
lithium abundances, we show in Figures 1 $\&$ 2 a series of
H-R diagrams for stars in narrow intervals of [Fe/H].
The lithium abundances span at least two dex from a maximum
of about $\log\epsilon$(Li) = 3.0 
 to upper limits of $\log\epsilon$(Li) $<$ 1.0.
On these two figures, the size of the symbol denotes the
lithium abundance; measured abundances are shown by open symbols and
limits by the filled symbols. Girardi et al.'s evolutionary tracks are plotted.
An alternative representation of the data is shown in Figure 3
where we plot the lithium abundances against the inferred stellar
mass with each panel again showing stars within a narrow range in [Fe/H].

\begin{figure*}
\includegraphics{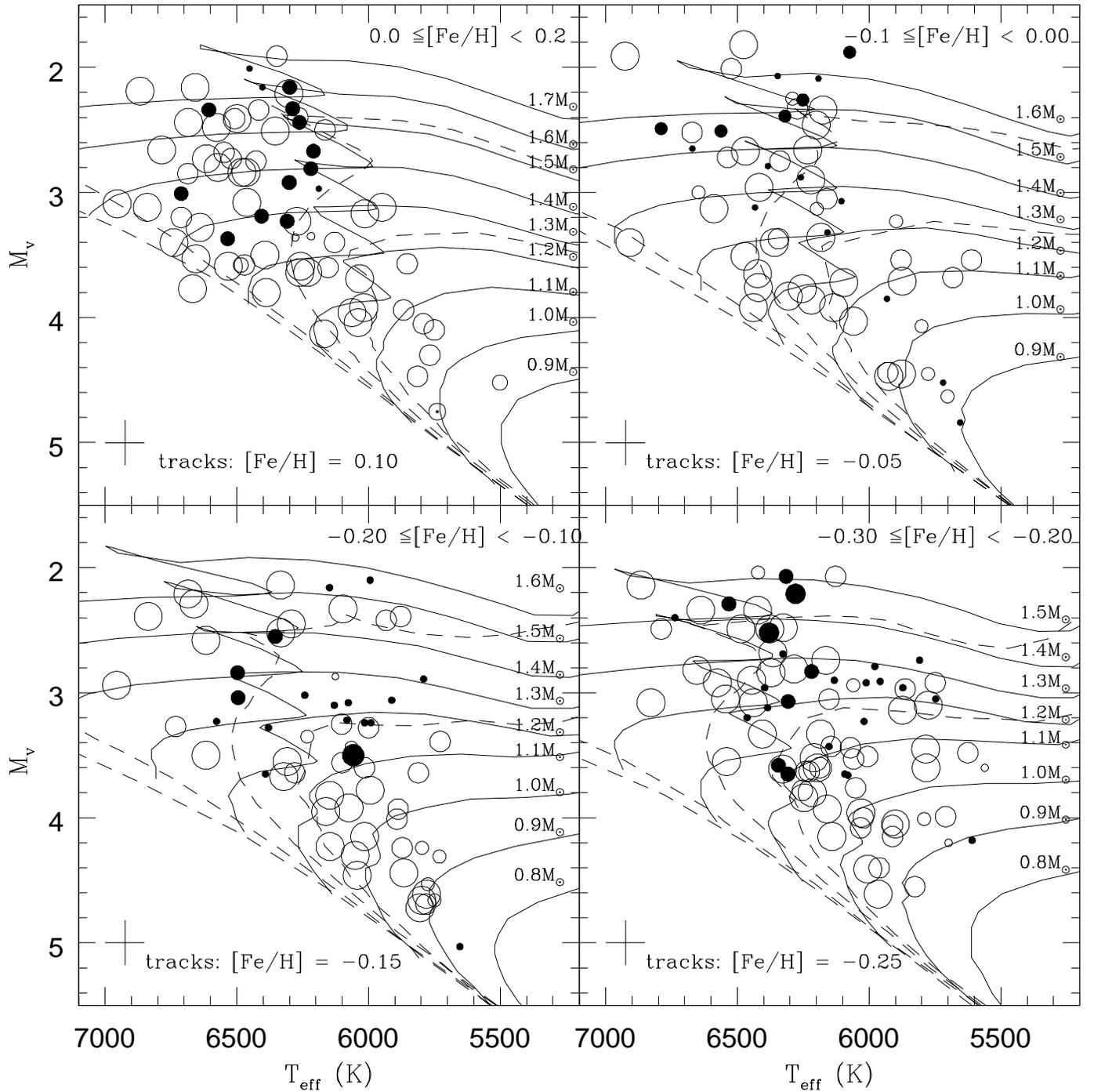}
\caption{H-R diagrams showing the stars from Table 1 for the
top four metallicity bins. Evolutionary tracks (solid lines),
and isochrones (broken lines) for 100, 700, 2500, and
4500 Myrs from Girardi et al.
(2000) are shown for the mean [Fe/H] of a bin. Filled circles
denote stars with  no detected lithium,  and open circles denote
stars with detectable lithium. 
The symbol size represents the magnitude of the
lithium abundance (or upper limit). Four sizes are used. In order of
decreasing symbol size, the abundance intervals are  
: log $\epsilon$(Li) $\geq$ 2.50,
2.0 $\leq$ log $\epsilon$(Li) $<$ 2.5, 
1.5 $\leq$ log $\epsilon$(Li) $<$ 2.0, and
log $\epsilon$(Li) $<$ 1.5.}
\end{figure*}

\begin{figure*}
\includegraphics{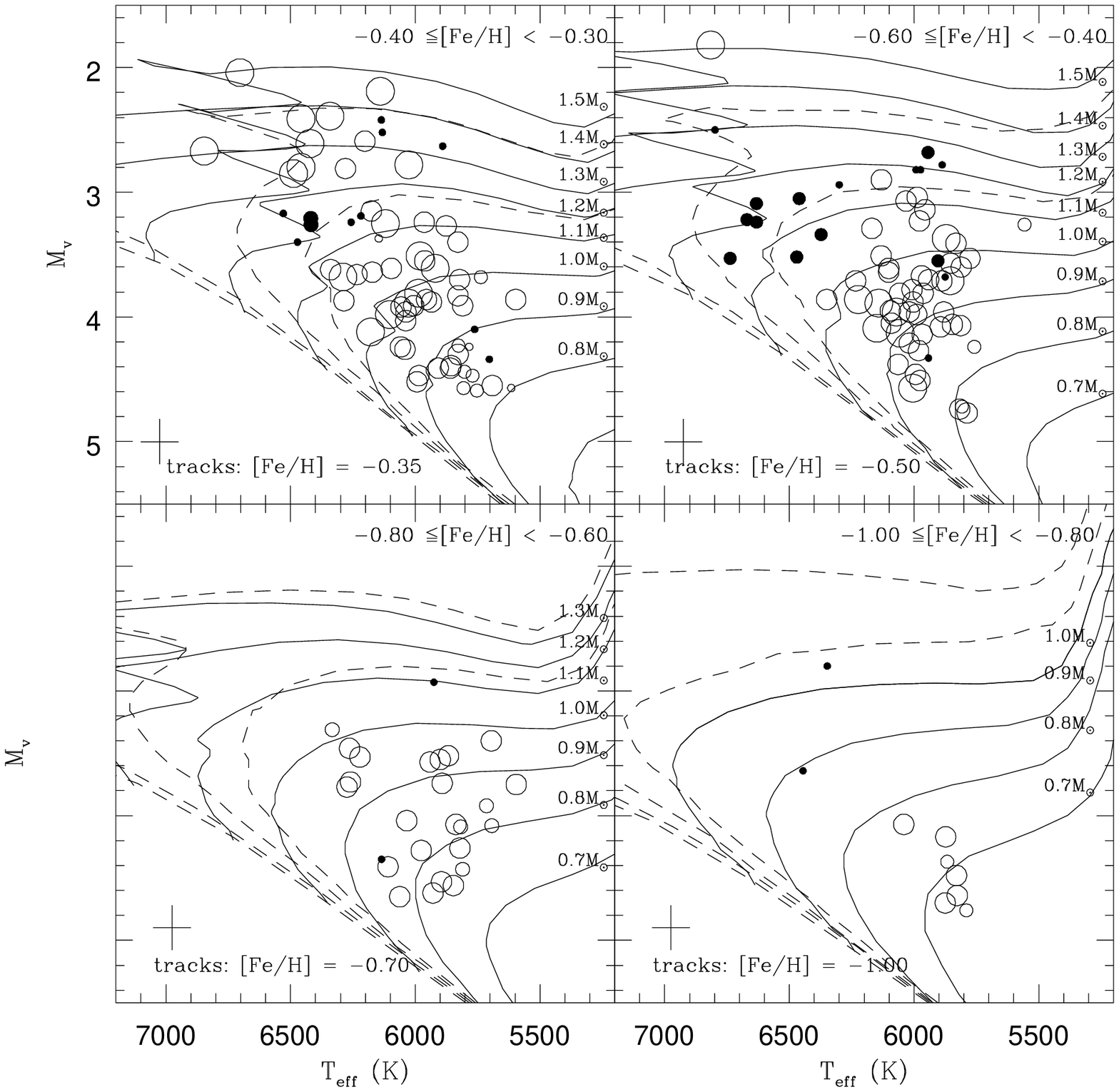}
\caption{H-R diagrams showing stars from Table 1 for the four
bottom [Fe/H] bins.
See caption to Figure 1 for further explanation.}
\end{figure*}

Before discussing our sample of   field
stars, we look at
 available data on lithium abundances  for main
sequence F-G stars in open clusters in the solar neighbourhood.
There is now a quite considerable collection of clusters
for which lithium abundances have been measured for F-G-K 
  stars from clusters with pre-main sequence stars to
clusters older than the age of the Sun.
In a review of embedded clusters in
molecular clouds, Lada \& Lada (2003) suggest that most field
stars originate from disrupted embedded clusters, and that
open clusters are the small fraction of surviving embedded clusters.
We supplement this suggestion with the assumption that 
field and cluster stars having
 the same basic parameters (mass, age, and composition) share a
common initial lithium abundance.
Recent reviews on lithium abundances of stars in open clusters
 include those by  Deliyannis (2000),
Jeffries (2000), and Pasquini (2000).

Unfortunately, the available cluster data  do not sample the
full range of metallicities and ages of the field stars. 
 The ([Fe/H],$t$) plane is sparsely
covered by the clusters.
 With one exception,
the open clusters for which lithium abundances have been reported
 cover the narrow range in [Fe/H] from  $+0.2$ to $-0.2$, ages 
from the pre-main sequence  to about 5 Gy, and are
  at the Sun's Galactocentric distance within about 0.5 kpc. The
exception is the metal-poor ([Fe/H] = $-0.5$; Hill \& Pasquini 2000)
 cluster NGC 2243 at about 2 kpc in
the anticentre direction.  This is not a representative of the
solar neighbourhood.
  We have taken lithium abundances
directly from the published papers on clusters and made no attempt to
make a uniform reanalysis of the observations. Use of different
temperature scales and consideration or neglect of Non-LTE
effects may have introduced systematic differences of, perhaps,
0.2 - 0.3 dex between different papers.  

\begin{figure*}
\includegraphics{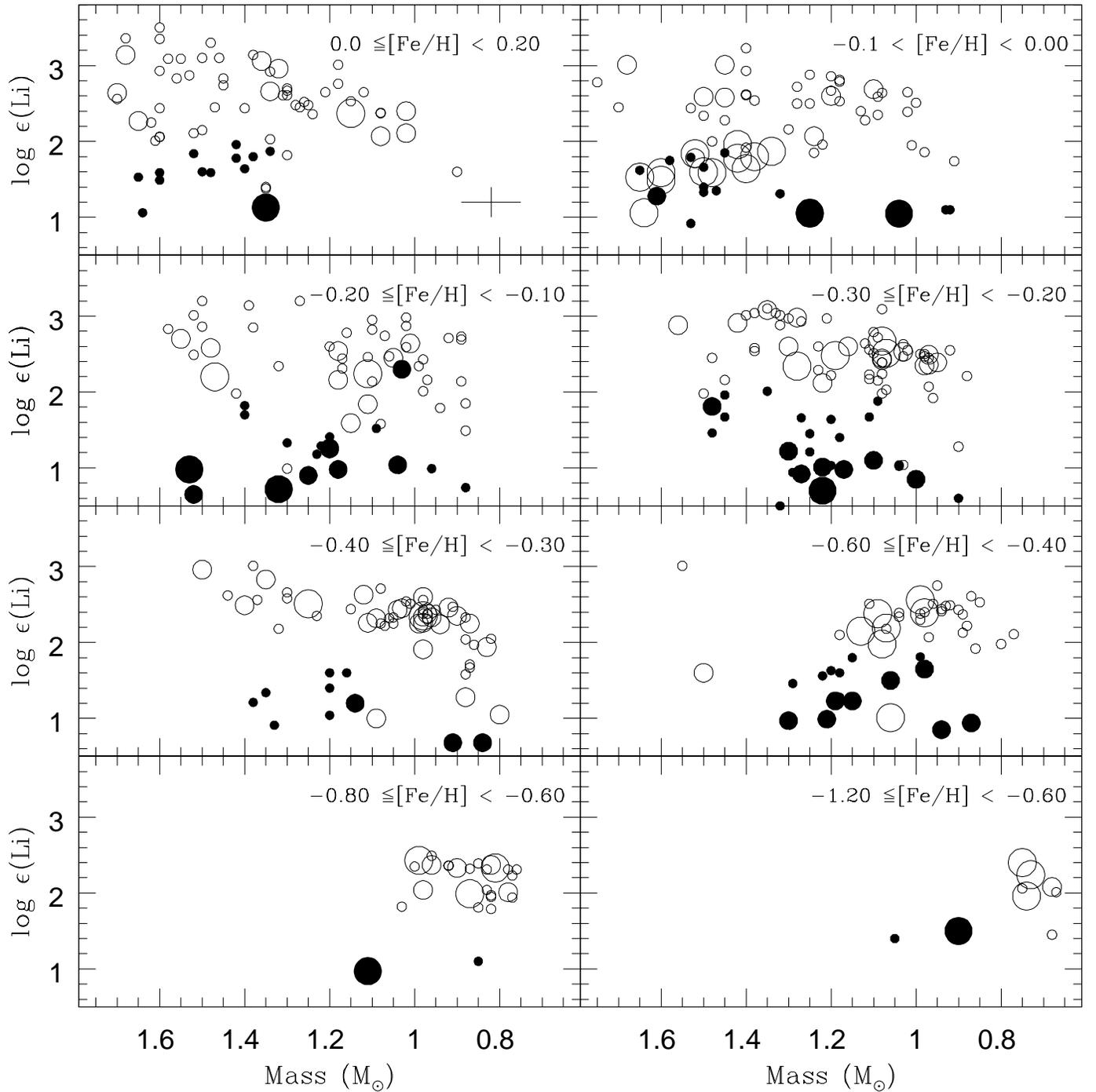}
\caption{Li abundance versus stellar mass in eight different
[Fe/H] bins. 
Filled circles
denote upper limits to the lithium abundance. Open circles refer to
measured abundances. The three sizes for the circles identify the
accuracy of the {\it Hipparcos} parallax, a major influence on the
derived mass. Stars with a  parallax error    
 $\geq$ 10$\%$ are assigned the biggest circle. Stars with an error of
5$\%$ to 10$\%$  are shown with the middle-sized circle.
Those stars with a  5$\%$ or smaller error are given the smallest
circles. }
\end{figure*}

\subsubsection{Initial lithium abundance at [Fe/H] $\sim$ 0}

In the youngest clusters, the lithium abundance is  constant 
 down to a mass of slightly less than 1$M_\odot$. The sample of such
 clusters includes these
with pre-main sequence stars: NGC 2264 (King 1998;
Soderblom et al. 1999) with $\log\epsilon$(Li) = 3.2, and the Orion Association
(King 1993; Cunha, Smith, \& Lambert 1995) also with
$\log\epsilon$(Li) = 3.2. Young clusters with ages of about 40 to 100 My
show a gradual decrease in lithium abundance with decreasing mass beginning at
about the solar mass. The lithium abundance of the stars more massive than a
solar mass is independent of mass and
 the same in each cluster: IC 2602 and IC 2391 (Randich
et al. 2001), IC 4665 (Mart\'{\i}n \& Montes 1997),
 $\alpha$ Persei (Balachandran, Lambert, \& Stauffer 1988, 1996;
Boesgaard, Budge, \& Ramsay 1988),  Blanco 1 (Jeffries \& James 1999),
and the Pleiades (Boesgaard, Budge, \& Ramsay 1988; Jones et al. 1996).
This abundance is in the range 
$\log\epsilon$(Li) = 3.0 to  3.2. 

 A remarkable, unanticipated, and  unexplained illustration of
astration is the Li-dip discovered by Boesgaard \& Tripicco (1986a)
among main sequence stars of the Hyades cluster. Cluster stars 
in a narrow temperature range centered on $T_{\rm eff}$ = 6600~K
show lithium depletions of up to 2.0 dex relative to 
stars  with $T_{\rm eff}$ $>$ 6400~K and $T_{\rm eff}$ $<$ 6200~K.
The Li-dip discovered in the Hyades cluster of
age $t$ = 660 My is  detected among members of the U Ma group at
$t$ = 300 My (Boesgaard, Budge, \& Burck 1988; Soderblom et al. 1993)
and M 34 at $t$ = 200 My (Jones et al. 1997). An  absence of the
dip in NGC 2516 at $t$ = 150 My (Jeffries, James, \& Thurston 1998),
M 35 at 175 My (Barrado y Navascu\'{e}s, Deliyannis, \& Stauffer
2001), and
NGC 6475 at $t$ = 220 My (James \& Jeffries 1997; Sestito et al. 2003)
possibly indicates that the dip is not present for stars younger than
about 200 My but it has to be noted that the absence
 may be a consequence of 
inadequate or no coverage of the $T_{\rm eff}$ range of the dip. The lithium
abundance on the high temperature side of the Li-dip is indistinguishable
from  the abundance seen in the youngest clusters: $\log\epsilon$(Li) = 2.9 (NGC 2516),
2.9 (M 34), 2.8 (M 35), 2.9 (NGC 6475), and 3.0 (U Ma). On the hot
side of the Li-dip there may be a star-to-star scatter in the Li
abundances which may cause estimates of a cluster's initial Li
abundance to be lower than the true value -- see our later
discussion of the field stars, especially Figure 4. 

In clusters with ages from about 300 My to about 2 Gy,
a signature of the Li-dip is present.
 No single cluster provides
an adequate  number of stars to affirm that the lithium abundance  of the hot
stars is independent of mass. This abundance seems, however, to be
very similar from one cluster
to the next and identical to within the errors with that from the
youngest clusters: $\log\epsilon$(Li) = 3.1 (Hyades -
 Boesgaard \& Tripicco 1986a; 
Thorburn et al. 1993; Balachandran 1995),
 3.2 (Praesepe - Boesgaard \& Budge 1988;
Soderblom et al. 1993; Balachandran 1995);
 3.0 (NGC 6633 - Jeffries 1997; Jeffries et al. 2002),
3.1 (NGC 752 - Hobbs \& Pilachowski 1986; Pilachowski \& Hobbs 1988; Balachandran
1995), 3.2 (NGC 3680 - Pasquini, Randich, \& Pallavicini 2001).
Observations of the Coma Berenices cluster (age of 400-500 Myr)
 provide just one star
on the hot side of the Li-dip (Ford et al. 2001). This star
with the abundance $\log\epsilon$(Li) = 2.8 may be on the
dip's shoulder.
The similar maximum lithium abundance in the youngest clusters and in these
slightly older  clusters implies that F main sequence stars experience
very little depletion of lithium  up to ages of about 1 Gy. The G and K stars
do experience astration which is more severe the lower the mass of the
star. 
Previous lithium detectives have noted the uniformity of the maximum lithium abundance
in local open clusters  and suggested that it represents the
interstellar lithium abundance of the clusters' natal clouds (e.g., Pilachowski
\& Hobbs 1988).
 The  uniformity implies little to no
evolution of the lithium abundance in the local
 interstellar medium over the last
several Gy. 
The interstellar Li abundance is possibly slightly less than the
solar system abundance ($\log\epsilon$(Li) = 3.3), a difference
which may reflect a systematic error in the lithium abundance
analysis of early F stars (i.e., classical atmospheres ignore stellar
granulation).

\subsubsection{The Li-dip at [Fe/H] $\sim$ 0}

 As noted above, the dip
is seen only in clusters older than about 300 My. In clusters older than
about 2-3 Gy, the dip is not seen among the main sequence stars
because stars susceptible to the responsible processes have evolved off the
main sequence. The few well observed clusters between these age limits
show the Li-dip. 
(In our view, the Hyades Li-dip is the sole convincing example
of a dip with well defined steep sides.  Other clusters clearly
have Li-poor stars at the temperature of the Hyades dip but the 
$T_{\rm eff}$-dependence of a dip cannot be defined clearly.)

A uniform analysis of observations of the Li-dip was reported by
Balachandran (1995) for the Hyades, Praesepe, NGC 752, and M 67. (In the
case of M 67, the Li-dip is not a main sequence phenomenon but is 
inferred from lithium abundances of subgiants.) A  conclusion
from this work is that over the limited metallicity range 
represented by the Hyades, Praesepe, and NGC 752 ($+0.12$ to $-0.15$ in [Fe/H],
according to Balachandran), the dip occurs at the same
effective temperature for the zero age main sequence stars ($T_{\rm eff}
\simeq 6500$K) which implies that the corresponding mass is
metallicity dependent ($M/M_\odot \sim$ 1.3 + 0.5[Fe/H]). Among additional
conclusions drawn by Balachandran are the following: lithium in stars of the
dip is destroyed within the stars; all main sequence stars within the temperature
range that defines the Li-dip  experience  astration. These
conclusions and the mass dependence are potentially examinable using samples
of field stars. In particular,  the greater [Fe/H] range covered
by field stars may allow for mass-metallicity relation to be
extended to lower [Fe/H]. 
Balachandran's  conclusions about the shape of the Li-dip are
likely too subtle to be checked using field stars.

\subsubsection{General astration at [Fe/H] $\sim$ 0}

The  Hyades stars  show
a monotonic decrease of lithium abundance for
masses below the dip, and no star-to-star scatter in abundance
at a fixed mass.  
The few well observed clusters of similar or older age to
the Hyades show
a lithium abundance versus mass relation not remarkably
different from the Hyades. Balachandran (1995) compares
Hyades with Praesepe (also 600 -700 My), NGC 752 (1100 My),
and M 67 (4000 My).  Randich et al. (2003) compare Hyades and
NGC 188 (7000 My). At fixed mass rather than a common $T_{\rm eff}$,
the run of the maximum lithium abundance with
mass is very similar across this age range, as noted by
Randich et al. who, however, use $T_{\rm eff}$ rather than mass for the comparison.
  On the assumption that
the initial lithium abundances were very nearly alike and the 
 assumption that astration is
composition-independent, this implies that the additional astration is
slight  above ages of about 600 My.
It is well known that  astration from birth to
the age of the Hyades is significant - approximately 1.0 dex
at 1$M_\odot$ (see above cited reviews). 
The growth of the astration--mass relation to the levels
 shown by the Hyades and older clusters 
is traceable from the cluster analyses cited above.
 Since low mass stars younger than the Hyades are
not  well represented in our sample, we refrain from commenting  further
on
these younger clusters, the growth of the lithium-mass
relation for low mass stars, and the scatter in lithium
abundances at a fixed mass.

Scatter in lithium abundances exists among low mass stars in
at least one old cluster.
Main sequence stars of M 67 show a star-to-star scatter of
at least 1 dex (Jones et al. 1999 and references therein) with about
two-thirds showing a (maximum) lithium abundance consistent with that of
other clusters (see above) and the  majority of the other
one-third showing no detectable lithium or an abundance limit about 1
dex below the lithium-rich stars. In sharp contrast is NGC 188 for which
Randich et al. (2003) find `virtually no scatter' in the lithium
abundance of solar-type stars, and an abundance consistent with that
of other old clusters (and the Hyades) including that of M 67's  Li-rich two-thirds.
 Randich
et al. further remark that `M 67 remains so far the only old cluster
for which a dispersion [in lithium abundances] among solar-type stars
has been confirmed', but the fact is that few old clusters have been
observed for lithium and in those few the sample of examined stars
is small.

A curiosity
is that the Sun's lithium abundance ($\log\epsilon$(Li) = 1.0 --
M\"{u}ller, Peytremann, \& de la Reza 1975)  appears to fall  by
more than 1 dex 
below  the trend defined by the field stars 
(see Figure 3). If placed among NGC 188's stars, the Sun
would be deemed very Li-poor. Among M 67's stars, the Sun would be
one of the most Li-poor stars.
 This hint that the Sun may be `peculiar' as regards the depletion of
lithium weakens its value as a calibrator for prescriptions of non-standard
modes of lithium astration.

\section{Lithium abundances of the field stars}

The field star sample is sorted into  eight  metallicity bins.
 In the H-R diagrams (Figures  1 and 2) for the eight bins, the size of the
open circle denotes the magnitude of the measured lithium abundance
and  upper limits are represented by a filled circle of the appropriate
size. 
The lithium abundance versus  mass plots are shown in 
Figure 3  with the size of the symbol denoting the accuracy of the
{\it Hipparcos} parallax, again open circles refer to measured lithium
abundances and filled circles to upper limits.
The parallax error is the principal contributor to the precision
with which the stellar mass may be estimated from a given set of
evolutionary tracks.

Quite independently of the lithium abundance, one fact stands out
from inspection of the HR-diagrams (Figures 1 and 2) and the
 lithium-mass plots (Figure 3): the absence of
the higher mass stars in the lower [Fe/H] bins. Contrast, for
example, the panels in Figure 3 for $-0.4 <$ [Fe/H] $> -0.3$ and
$-0.6 <$ [Fe/H] $-0.4$.
The former panel is representative of the higher
[Fe/H] bins:
 the maximum mass in a bin   increases 
with increasing [Fe/H]. (A mass of about 1.6$M_\odot$ is about the
mass at which a main sequence star is so hot that the Li\,{\sc i}
resonance doublet is undetectable. Lithium is traceable in higher
mass stars when, as subgiants, they evolve across the Hertzsprung
gap as subgiants.)
 The latter panel is almost devoid of stars
with $M >1.2M_\odot$, and the maximum mass seen in the panels
decreases with decreasing [Fe/H].
 This correlation  probably reflects the form of the
age-metallicity relation displayed by the field stars (Edvardsson et al.
1993; Chen et al. 2000; Reddy et al. 2003).  

The age-metallicity relation for local field stars,
 as first clearly expressed by Edvardsson et al., exhibits 
scatter at a fixed age. In the (age, [Fe/H]) plane, the
stars are bounded at the upper end at [Fe/H] $\sim +0.2$
at all ages, but at the lower end by a trend of decreasing
[Fe/H] with increasing age. For example, at [Fe/H] = $-0.5$, there
are very few field stars (at least, in these surveys) younger than
about 4 Gy, and, as a result, the H-R diagrams and the
low-[Fe/H] bins in Figure 3 lack the more massive stars. 
This interpretation is independent of the origins of the scatter
in the age-metallicity relation. A search for field stars more massive
than those represented in the Figure 3 would be of interest.

\begin{table}
\centering
\caption{ Mean Li abundances calculated from the six most Li-rich
stars in each [Fe/H] bin.}
\begin{tabular}{@{}rrc@{}}
\hline \hline

[Fe/H]    &      log $\epsilon$(Li)   &  $< m/m_{\odot} >$  \\
\hline

0.2 to \phantom{$-$}0.0   & 3.33$\pm$0.16 &     1.52$\pm$0.12  \\
\phantom{-}0.0 to $-$0.1 & 3.03$\pm$0.14 &    1.40$\pm$0.16  \\
$-$0.1 to $-$0.2 & 3.10$\pm$0.06 &   1.31$\pm$0.17\\
$-$0.2 to $-$0.3 & 3.09$\pm$0.04 &  1.31$\pm$0.11\\
$-$0.3 to $-$0.4 & 2.86$\pm$0.14 & 1.34$\pm$0.14\\
$-$0.4 to $-$0.6 & 2.64$\pm$0.07 & 1.03$\pm$0.26 \\
$-$0.6 to $-$0.8 & 2.39$\pm$0.06 & 0.91$\pm$0.06 \\
$-$0.8 to $-$1.2 & 2.15$\pm$0.15 &  0.71$\pm$0.04 \\
\hline
\end{tabular}
\end{table}

\subsection{The maximum lithium abundance}

The maximum lithium abundance  in each [Fe/H] bin is taken to be
the mean of the six highest lithium abundances in that bin.
Table 2 shows this
abundance and the corresponding mean mass.
Selected stars are members of the thin disk. Only in the lowest
metallicity bin would  thick disk stars  be members of the
richest sextet. The two thin disk stars in this
bin have abundances of $\log\epsilon$(Li) = 2.40 and 1.96 for
a mean of 2.18, a value equivalent to the mean of 2.15 from
the entire sextet.
For the three  bins which overlap the [Fe/H] range of the clusters,
the lithium abundances in Table 2 are identical to the maximum
abundances from the clusters. We identify this abundance with that
of the interstellar gas from which the clusters and field stars
formed. The lithium abundance seems to have increased by only
about 0.2 dex as [Fe/H] increased from $-$0.35 to $+$0.15.

At metallicities [Fe/H] $< -0.4$, the maximum lithium
abundance is provided by stars of about 1$M_\odot$ or less, a
lower mass than provides the maximum abundance 
for more metal-rich stars. Astration of lithium in the latter stars
seems unlikely from Figure 3 and the data on open clusters
discussed above. But,  the question
arises: is the maximum lithium abundance for the [Fe/H] $< -0.4$ bins
reduced from the initial abundance by astration? 
 Inspection
of the data in the  $-0.4 <$ [Fe/H] $ -0.3$ bin
shows the lithium abundance falling
steadily with decreasing mass. If a similar relation, as seems
plausible, applies to the
adjacent bin for $-0.6 <$ [Fe/H] $< -0.4$, the maximum
abundance attributed in Table 2  to the bin must be increased by about
0.5 dex, as indicated by the outlier at $M \simeq 1.55M_\odot$. Then, the
entry in Table 2 for the bin would be $\log\epsilon$(Li) $\simeq 3.1$ 
leading to the implication that the lithium abundance was constant
as the thin disk's [Fe/H] increased from $-$0.5 to $+$0.1.

If astration is ineffective in the contributing stars,
 the lithium abundances in Table 2
map the evolution of lithium with [Fe/H]. 
The orthodox consider that the floor to the tabulated data
 be set at the lithium abundance ($\log\epsilon$(Li) $\simeq 2$)
of the Spites' plateau of the halo stars (Spite \& Spite 1982).
An allowance for astration, as suggested by Figure 3 and the bins for
[Fe/H] $> -0.4$, would seem to imply that the the lithium abundance
has evolved little over the lifetime of the thin disk. Then, how should one
match a near uniform abundance of  $\log\epsilon$(Li) $\simeq 3.1$ for the
thin disk to the halo abundance of  $\log\epsilon$(Li) $\simeq 2$.
Perhaps,  one should contemplate models in which the
gas providing the earliest generations of thin disk stars
enriched  in lithium
from sources other than the Big Bang. 
 The unorthodox view of the Spites' plateau is that the Big Bang
lithium abundance was greater than the present plateau; lithium has been depleted
over the life of the halo stars defining the plateau.
This view encourages the speculation
that the disk's lithium abundance may have evolved only slightly above
the Big Bang value. 
In this connection, we note that a standard big Bang model with the
$\Omega_{\rm b}$ estimated from the CMB fluctuations and the
cosmological D/H ratio implies a lithium
abundance of $\log\epsilon$(Li) $\simeq 2.6$ (Kirkman et al.
2003), a value which would
merge smoothly with the entries in Table\,2 for the more metal-rich bins.
This scenario supposes that lithium is astrated in the low mass
metal-poor stars comprising the Spites' plateau and, by extension, also
in the stars contributing to the entries for the lowest two
bins in Table 2.

In the simplest of thin disk models, one expects the lithium abundance
to increase with time. The data for the individual stars contributing to
Table 2 hint at such an increase, but we suspect that the biases in
the samples comprising the three surveys are largely responsible for the
trend, also seen in [Fe/H] -- age trend. The abundance ratio Li/Fe is
remarkably constant across the bins: $\log\epsilon$(Li/Fe) = $-4.30\pm$0.10,
$-4.45\pm$0.11, $-4.24\pm$0.06, $-4.20\pm$0.06, $-4.35\pm$0.18, $-4.32\pm$0.16,
$-4.44\pm$0.04, and $-4.41\pm0.27$ for the bins from the most iron rich to
the most iron deficient.

  Previously, the evolution of lithium with
[Fe/H] has been identified with the upper envelope of the points in
a $\log\epsilon$(Li) versus [Fe/H] plot (see, for example, Rebolo et al. 1988;
Lambert et al. 1991) without specific consideration of the mass of
the stars and the possible variation of the mean mass with [Fe/H] and its implications
for astration. Our
discussion based on Figure 3 draws attention to this variation.

\subsection{Star-to-star scatter?}

If initial lithium abundance were coupled tightly to the  [Fe/H] of
a star,
astration at a fixed [Fe/H] were   only a function of a star's
mass and  were completed before evolution off the
main sequence,  the lithium abundances in each panel of Figure 3
would be a smoothly varying function of mass with no scatter
about the mean trend except for that attributable to errors of
mneasurement. This is not the
case, an observation noted previously by Balachandran (1990) and Chen et al.
(2001) from their samples of stars observed off the zero-age
main sequence.  

\subsubsection{Field and cluster stars at [Fe/H] $\geq -0.3$}

We begin examination of the field stars with
the 0.0 $\leq$ [Fe/H] $< +0.20$ bin and the Hyades stars.
The comparison is shown
in Figure 4.
The solid line is a mean line through the Hyades' points
taken from Balachandran (1995).
The field stars at  masses less than those of the Li-dip
 trace the Hyades'
mean line with few stars showing a lower lithium abundance.
The low-mass limit (red-edge)  of the Hyades' Li-dip appears to mark the
low-mass boundary of the large star-to-star scatter in Figure 4.
 Our
sample, however, contains few stars with $M < 1M_\odot$. Published
surveys providing Li abundances for such stars indicate that 
scatter in lithium abundances exists for such stars (e.g.,
Pasquini, Liu, \& Pallavicini 1994, Malik, Parthasarathy, \& Pati 2003).

The  field stars mapping the Hyades relation are much older than
the cluster: the mean age of the field sample on the low mass side
of the Li-dip is about 5 Gy versus the
cluster's age of 660 My. At face value, the similar lithium abundances
of field and Hyades stars of the same mass on the low mass side of the
Li-dip
 would seem to confirm the suggestion from
cluster analyses that the rate of astration is very slow in
stars older than about the Hyades. There is, however, a systematic
slight difference in the mean [Fe/H] of the field stars and the Hyades:
[Fe/H] for the field is about 0.07 dex less than that of the cluster.
Astration  is predicted to be less severe in lower [Fe/H]
stars (Chaboyer et al. 1995), and, hence, the field stars
appear to have undergone more severe astration than the
Hyades, say by about 0.3 dex in about 4 Gy. This 
suggestion is not inconsistent with the indications from
comparisons of the Hyades and older clusters.

Lithium abundances in field stars with masses greater than the value
corresponding to the red edge of the Li-dip show a range  from
a star's initial abundance ($\log\epsilon$(Li) $\simeq$ 3.0) to the
detection limit for these warm stars, a limit of $\log\epsilon$(Li) $\leq$ 1.5
or even less. The fact that the lower mass limit for severe
scatter in the field sample coincides with the red edge of the
Hyades Li-dip is presumably not fortuitous but a reflection of a common
cause.
It is a well known observation (Kraft 1987) and one that
motivated Balachandran's selection of field stars that 
 early-F stars include rapid rotators and the break in
rotational speeds occurs at about the mass of stars
belonging to the Li-dip. We follow Balachandran in supposing that
loss of angular momentum (`braking') has led to
mixing and destruction of lithium.
Given an initial distribution in angular momenta, it is easy
to imagine how a spread in surface lithium
abundances results after braking.
Clearly, the Hyades Li-dip is not simply replicated by the
field stars. Field stars of normal lithium abundance occupy the
mass range of the dip. Li-poor field stars are found at masses higher than
the central mass of the Hyades dip. 
The filling in of the Hyades dip by field stars may be
due to  errors in the assigned masses arising primarily from parallax and
$T_{\rm eff}$  errors and the spread in the central mass of the dip resulting from
its metallicity dependence and the 0.2 dex width in [Fe/H] of the bin.
 
Young well-sampled clusters (e.g., $\alpha$ Per, and Pleiades)
 show neither the Li-dip nor a spread in
lithium abundances for stars with masses greater than the mass of the
subsequent red-edge of the Li-dip. Clusters with a clear signature of the
Li-dip have very few stars more massive than those belonging to the Li-dip,
and generally few stars in the dip. Even the Hyades could be suspected
of showing some scatter in lithium abundance in excess of observational errors  
for stars more massive than the mass of the centre of the dip. We venture to
suggest that the large scatter shown by field stars is not
contradicted by observations of open clusters. Given that the field
stars are older than the cluster main sequence stars of the same
mass and that the timescale for development of lithium astration may
be comparable to or longer than the main sequence lifetime, it is
possible to understand the possibly smaller scatter of the lithium
abundances among  the cluster.

Our diagnosis of Figure 4 seems to apply to  the other bins (Figure 3)
 down to
[Fe/H] = $-0.3$ with the proviso that the central mass of the Li-dip
moves to lower values with decreasing [Fe/H]. 
There is a suspicion that the $-0.2 \leq$ [Fe/H] $< -0.1$ bin shows
appreciable scatter in the lihium abundances among stars
to the low mass side of the Li-dip. 
At [Fe/H] $\sim -0.3$, the distribution of the data points in Figure 3
(also Figure 2) changes from  one dominated by scatter to one
in which a clear trend emerges with a few points lying on the
low lithium side of the trend.

\begin{figure*}
\includegraphics{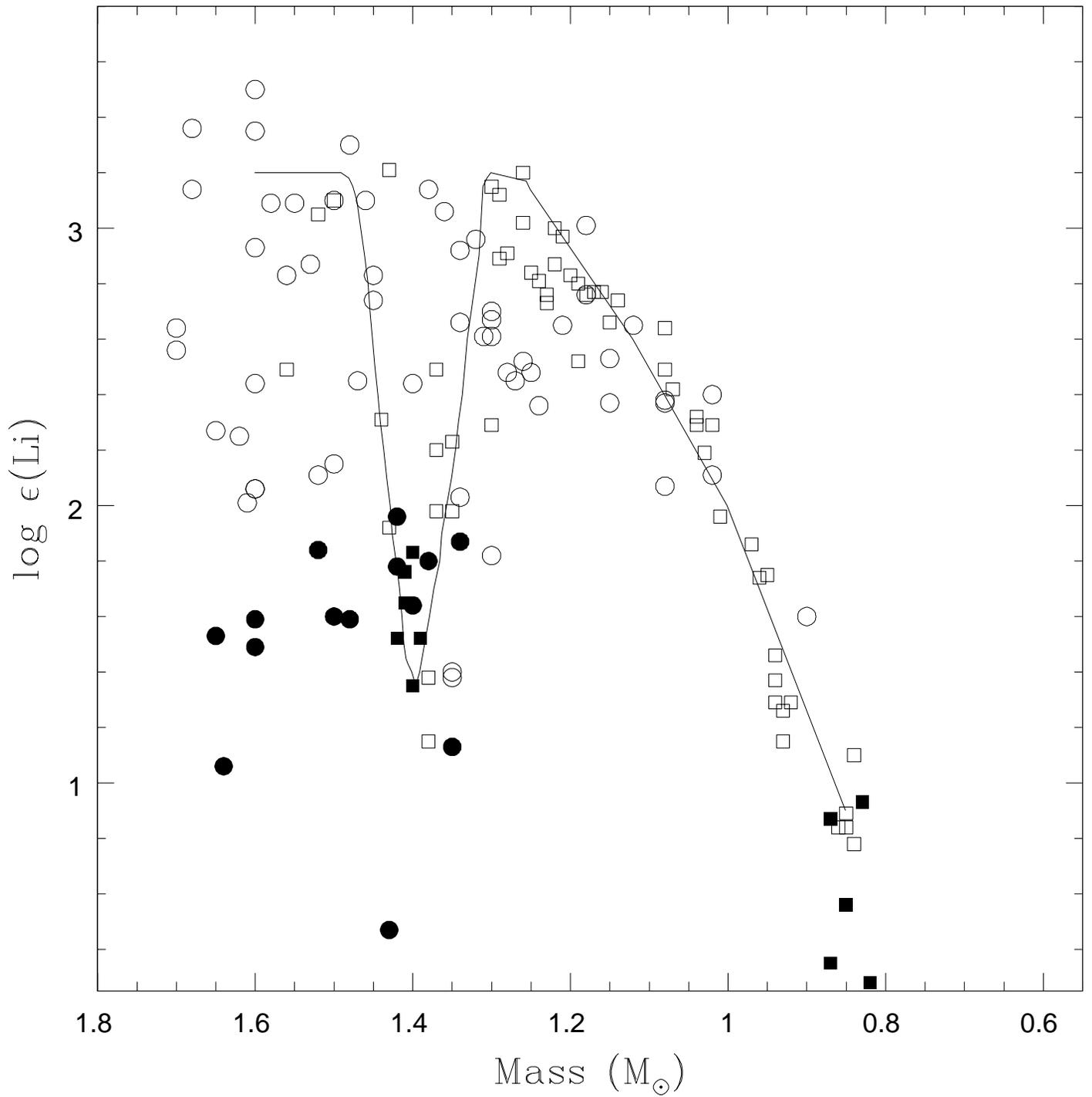}
\caption{Lithium abundance versus stellar mass for the Hyades
members (open squares) and field stars with [Fe/H] in the range 0.0 to $+0.2$
(circles). Filled symbols denote upper limits to the lithium abundance.
The solid line is the mean relation for the Hyades stars with the Li-dip
prominent at about $M = 1.4M_\odot$.}
\end{figure*}

In young clusters such as the Pleiades, a dependence of the lithium abundance
on rotational velocity ($v \sin i$) is seen for stars of less than
about 1$M_\odot$ with higher lithium abundance tied to faster rotation. This
dependence is not seen in the more massive Pleiades stars nor at
all strongly in the
low mass stars of very young clusters (e.g., $\alpha$ Per).
Also, the scatter in lithium versus mass relation is not
seen among the low mass Hyades stars. 
Assuming that the Pleiades-Hyades difference is not a reflection of different
initial conditions for the two clusters,
 it would seem   that  a different coupling of rotation and lithium 
 operates at and below
1$M_\odot$ than at the higher masses.

\subsubsection{The Li-dip}

Observations of the Li-dip in open clusters show that
over a narrow range in metallicity centred on the solar
value that the mass at which the Li-dip occurs decreases
with decreasing [Fe/H] (Balachandran 1995).
An alternative description of the
result is that the Li-dip occurs at a similar effective
temperature on the zero-age main sequence, which may be a clue to the
dip's origin.
The [Fe/H] dependence of the Li-dip 
was anticipated from samples of field stars by Balachandran (1990)
and echoed by Chen et al. (2001). 
 
The Li-dip is traceable in the H-R diagrams and the panels of
Figure 3. It is blurred on the high-mass
side by  scatter.
An occasional lithium-poor star of low mass (a solar
analog?) is readily set aside. We estimate by 
inspection the mass of the Li-dip by looking for the mass at which there
is a concentration of Li-poor stars and a relative paucity of
stars with a normal lithium abundance.  This exercise is impossible
for the bin with $0.0 <$ [Fe/H] $< +0.2$ and for bins with
[Fe/H] $< -0.6$. With imagination, the mass is about
1.35$M_\odot$, 1.25$M_\odot$, 1.2$M_\odot$, and 1.2$M_\odot$ for
the five bins between [Fe/H] = 0.0 and $-0.4$. In the bin for
[Fe/H] from $-0.4$ to $-0.6$, the low mass side of a dip seems traceable
and $M_\odot$ $\geq 1.2$ is suggested. Local clusters suggest
that the mass of the dip decreases by about 0.05$M_\odot$ for a
0.1 dex decrease in [Fe/H] (Balachandran 1995).
 This is not inconsistent with the
above estimates for the five bins whose  centres span 0.35 dex.

A key open
question remains: is the Li-dip preserved as the stars
age. Data on clusters are too sparse
to answer the question. Preservation of the dip would suggest that lithium
in Li-dip stars is destroyed
and not merely hidden below the surface.
 A signature of a preserved Li-dip 
would be a string of Li-poor stars along a track at
a constant mass (Figure 1 and 2). This is seen  for the bin with
$-0.6 <$ [Fe/H] $< -0.4$, but, in the more metal-rich bins,
 Li-poor  and Li-rich stars are commonly
juxtaposed.

 This juxtaposition may have several origins
and may not be in conflict with the idea that the Li-dip
is a permanent mark for a star until growth of the convective
envelope dilutes the surface lithium abundance of all stars. Errors in the
lithium abundance determination, the estimate of $M_V$, and the confusion
arising from the hook in the evolutionary track, 
 may all blur  evidence for a
Li-dip confined to and a permanent feature of main sequence
stars in a narrow mass range. Yet,  our assessment of these
effects is that they may not fully  account for the juxtaposition of Li-normal
and Li-poor stars.
 We suspect that  an additional
mechanism  is needed to account for the mixing of Li-rich and Li-poor
stars,

For [Fe/H] $< -0.3$, the distribution of the data  in Figure 3
(also Figure 2) changes from  one dominated by scatter to one
in which a clear trend emerges with a few stars on the
low lithium side of the trend.  A Li-dip seems apparent for the
bins with [Fe/H] between $-$0.3 and $-0.6$.

\subsubsection{Scatter at $M \leq 1M_\odot$}

An intriguing result from  open clusters is the contrast between
the dispersion of 1 dex or more
 in lithium abundances  among low mass stars of  the old cluster M 67 and
the uniformity of the abundances in a similarly old
cluster NGC 188. In light of this, we comment on
the scatter among the lithium abundances of the field stars
on the low mass side of the Li-dip.

A M 67-like scatter for low mass stars is seen in the bin covering [Fe/H] 
values from $-0.1$ to $-0.2$, and in the $-0.2$ to $-0.3$ bin.
Pasquini (2000) speculates that the scatter for M 67 is of
a bimodal nature but this is not evident from Randich et al.'s
(2003) presentation of the results (their Figure 4).
 A bimodality is not evident in the metal-rich bins
in Figure 3. 
  The scatter  appears 
smaller in the two more metal-rich bins 
(Figure 3), but we  note that the  number of
stars in these two bins
 with a mass of about one solar mass is small. The scatter
appears much reduced  for the  [Fe/H] $< -0.3$ bins. 
It is intriguing that the scatter at low
masses may be correlated with that among stars more massive than
the mass of the Li-dip. This hints at a common underlying origin for
the scatter -- a spread in angular momenta of main sequence
stars which is large at high [Fe/H] and reduced or absent in low
[Fe/H] stars? Our catalogue of field stars contains few stars
with $M \leq 1~M_\odot$. Pasquini et al. (1994) find a M67-like
scatter in this mass range from observations of field G stars.
Favata, Micela, \& Sciortino's  (1996)  results for nearby G and K
dwarfs may indicate a reduction of the scatter (see their Figure 1)
at about  $M \simeq 0.8M\odot$ and lower masses. These G and K
field stars are likely very old.

To search for an age dependence for astration,
  we selected
narrow well-sampled  mass intervals and plotted the lithium
abundance versus the age. In these samples, almost all of the stars
have evolved away from the zero-age main sequence. The spread in
ages across the sample is small relative to the total main
sequence lifetime. The lack of a dependence of lithium
abundance on age is not, therefore, in conflict with the observed
decrease of lithium with increasing age presented using the open 
clusters which span a larger age range than the field stars.

\section{Concluding remarks}

Unravelling the web of factors that control the surface
lithium abundance of main sequence F and G stars has
demanded extensive observations of lithium in stars in
open clusters and the field. Our principal goal in
this paper has been to present the first measurements of the lithium
abundances in nearly 200 F and G field stars, and to combine our results
with those of two earlier surveys (Balachandran 1990; Chen
et al. 2001) to provide a catalogue of lithium abundances
for 451 F and G field stars.

The field stars are presently residing in   the immediate neighbourhood of the
Sun, and most were likely born at about the Sun's Galactocentric
distance. Distance and, hence, absolute magnitude $M_V$ are
rather precisely known thanks to the {\it Hipparcos}
satellite. Since the three contributing samples include
stars somewhat off the zero-age main sequence, it is possible
with the theoretical evolutionary tracks and the measurements
of $M_V$, $T_{\rm eff}$, and [Fe/H] to assign an evolutionary
age to most stars.  The sample of open clusters for which
lithium abundances have been determined for F  and G
main sequence stars are also residents of the solar
neighbourhood, albeit  at a mean distance from the Sun somewhat
larger than the mean distance of the field stars. 
The environments in which field stars and open clusters
are formed are unlikely to have differed greatly and one
 expects a field star and an identical
cluster star to exhibit the same lithium abundance.

Lithium abundances of field and cluster
stars share  common aspects, and may point to some
differences. One common aspect is that the inferred
initial lithium abundance is the same over the age and
metallicity range spanned by the open clusters.
The initial or interstellar lithium
abundance has been approximately constant in the solar
neighbourhood for several Gyr and over the [Fe/H] range
from about $+0.2$ to $-0.2$. The growth of the lithium
abundance from the Spites' plateau to 
the present level is presently definable only through
observations of field stars. Table 2 is our attempt to
define that growth but it is based on the key assumption that
the lithium of the  defining stars is unaffected by astration.
If the astration resembled that exhibited by the Hyades and other clusters,
the lithium of the thin disk may have had an approximately
constant abundance as [Fe/H] grew from $-1$ to 0.

Astration of lithium is not merely a function of a star's mass,
age, and composition. This was evident long ago. The difference
between the low lithium abundance of the Sun and the much higher
abundance exhibited by many field and cluster solar-like stars
is one indication that other variables affect the surface lithium
abundance. There is probably no convincing evidence yet that
field and cluster stars of the same mass, age, and composition
differ in their star-to-star variation of the lithium 
abundances, a variation attributed to astration rather than a spread in
the initial lithium abundances.

 We have supposed that the
scatter in lithium abundances among field stars with masses
above the high-mass edge of the  Li-dip is due to differences
in initial angular momentum and its loss (rotational braking). 
An apparent absence of scatter in lithium abundance among  such stars
of open clusters may be due to the fact that
 few cluster stars  
in this mass range present themselves for analysis.
 It appears that the astration
responsible for the star-to-star scatter occurs in main sequence
not pre-main sequence stars.  

Stars on the low mass side of the Li-dip experience astration which
develops at an early age, i.e., it is evident in Hyades and younger
clusters. 
Cluster and field
stars show that additional astration beyond the age of the Hyades is
slight.
At [Fe/H] $\leq -0.2$, questions of lithium
synthesis and astration must be answered from 
field stars; few clusters with [Fe/H] $\leq -0.2$
 are known.  Only one (NGC 2243)  has been
examined for lithium (Hill \& Pasquini 2000). 
The field stars suggest that the star-to-star 
variation in lithium abundance is much reduced at
low metallicity with the exception of a few stars with
an uncommonly low lithium abundance. 

Despite intensive observing campaigns on open clusters and
field stars there remain gaps in our knowledge about lithium
abundances in the (mass, metallicity, age) space, and rotation
should be an additional one or two dimensions to this space.
Analyses of local  field stars should be extended by
making either more detailed studies of parts of the HR diagrams
sampled by the three surveys used here or by extensions to
new parts of the HR diagram, especially to lower masses.

We have
noted several times that the local open clusters span a
narrow range in [Fe/H]. To extend the work on clusters to
lower [Fe/H], it will be necessary to seek
distant clusters such as NGC 2243
 in the anticentre direction -- see, for example,
the catalog of open clusters compiled by Chen, Hou, \& Wang
(2003). Thorough scrutiny of a Hyades-like cluster of [Fe/H] $\sim$ $-0.5$
or less would offer an empirical test of the variation of the
mass dependence of astration with [Fe/H], a crucial test
for establishing the growth of the lithium abundance in the local
thin disk. The lack of metal-poor open clusters near the Sun and belonging to
the thin disk is presumably due to the  evaporation and dissolution of clusters.
Nonetheless, survival of old solar metallicity clusters like M 67 and NGC 188 and the
lack of (say) local [Fe/H] $\sim -0.5$ clusters seems odd. Certainly, these
clusters 
appear to exclude a rapid  recent increase in [Fe/H] to
present values.

\section{Acknowledgments}

We thank Neal J. Evans II for a helpful discussion.
We are indebted to Jocelyn Tomkin for obtaining the spectra
from  which our lithium abundances were derived. We thank referee
R.D. Jeffries for a constructive and helpful report.
This research has been supported in part by  
the Robert A. Welch Foundation of Houston, Texas.
This research has made use of the SIMBAD
data base, operated at CDS, Strasbourg, France, and the NASA ADS, USA.

{}

\end{document}